Andrzej Jarynowski[1,2,3]
Vitaly Belik[4]
[1]Smoluchowski Institute of Physics, Jagiellonian University in Cracow
[2]CIOP - National Research Institute and Military Institute of Hygiene and Epidemiology in Warsaw
[3]Interdisciplinary Research Institute in Wrocław
[4]System Modeling Group, Department of Veterinary Medicine, Freie Universität Berlin

# Modeling the ASF (African Swine Fever) spread till summer 2017 and risk assessment for Poland

**Abstract:** African Swine Fever (ASF) is viral infection which causes acute disease in domestic pigs and wild boar. Although the virus does not cause disease in humans, the impact it has on the economy, especially through trade and farming, is substantial. Recent rapid propagation of the (ASF) from East to West of Europe encouraged us to prepare risk assessment for Poland.

The early growth estimation can be easily done by matching incidence trajectory to the exponential function, resulting in the approximation of the force of infection. With these calculations the basic reproduction rate of the epidemic, the effective outbreaks detection and elimination times could be estimated.

In regression mode, 380 Polish counties (poviats) have been analysed, where 18 (located in Northeast Poland) have been affected (until August 2017) for spatial propagation (risk assessment for future). Mathematical model has been applied by taking into account: swine amount significance, disease vectors (wild boards) significance.

We use pseudo-gravitational models of short and long-range interactions referring to the socio-migratory behavior of wild boars and the pork production chain significance. Spatial modeling in a certain range of parameters proves the existence of a natural protective barrier within boarders of the 'Congress Poland'. The spread of the disease to the 'Greater Poland' should result in the accelerated outbreak of ASF production chain.

In the preliminary setup, we perform regression analysis, network outbreak investigation, early epidemic growth estimation and simulate landscape-based propagation.

**Problem:**

Recent rapid spread of the African Swine Fever (ASF) in the Northeast Poland during summer 2017 [Fig. 1] encourages us to prepare risk assessment for the whole country and predict future geographical transmission paths. We used available data on outbreak form OiE (World Animal Health Organization) [Fig. 1] and regional information from Statistics Poland (GUS) [Pic. 3]. African swine fever is viral infection which causes acute disease in domestic pigs and wild boar. Although the virus does not cause disease in humans, the impact it has on the economy, especially through trade and farming, is substantial. The occurrence of swine fever is associated with huge costs for pork producers whose share in GDP in Poland is estimated to be about one-third of national spending on science. As a consequence of ASF in Poland the value of pork and pig exports have been already reduced by hundreds millions of EURO. This project should and as a decision support system as a tool for epidemiologists and veterinarians in chief (Mur, et al. 2012).

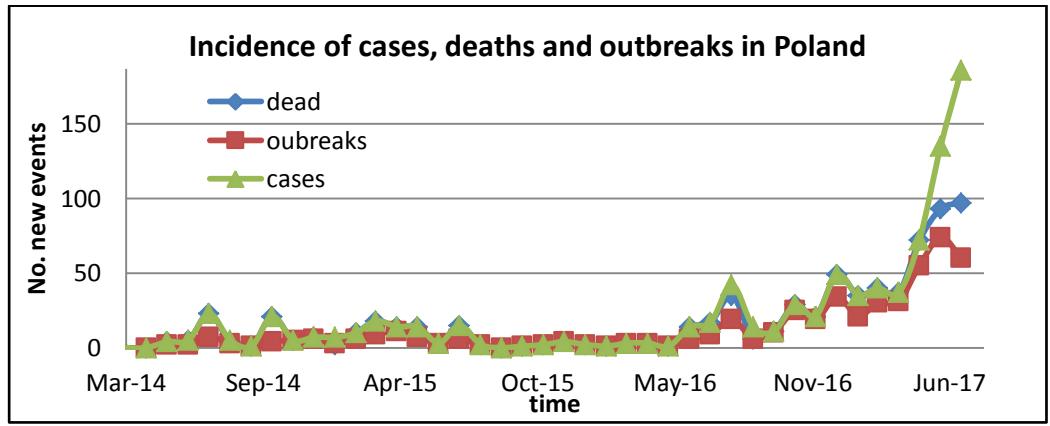

Fig. 1) Monthly incidence rates of ASF in Poland until August 2017 (Own figure based on data from OiE).

The disease has been occurring in Poland since February 2014, however, only in 2017 it exceeded the so-called population reproduction number ($R_0$) of the epidemic (adjusted epidemic reproduction rate). The time is running and most of Poland is currently at the highest risk of becoming endemic. Although the Polish Veterinary Inspection with and National Research Institutes are monitoring and controlling the disease spread, a mathematical model of the spatial and temporal dynamics of the disease spread is missing.

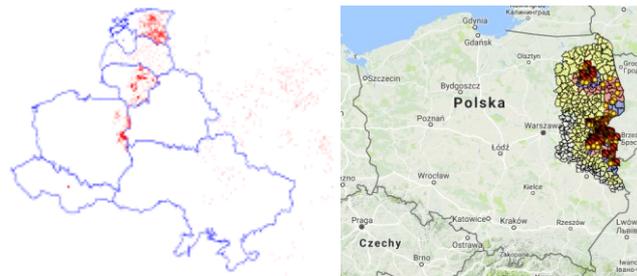

Pic. 1) [Left] Registered ASF cases in Europe up to August 2017 (Own map based on data from OiE). [Right] Registered ASF cases in Poland in 2017 up to August 2017 (source www.wetgiw.gov.pl)

We focus on a predictive stochastic ASF model based on empirical geographical data incorporating organizational network of regions, empirical forest, swine, and wild boar density as well as theoretical organizational structure of the pork production supply chain. This model would be equipped with decision support systems as a tool for epidemiologists. In the preliminary setup, we perform early epidemic growth estimation and simulate landscape-based propagation.

**Preliminary Results - Early detection:**
Set of dependent variables stocks counts farm/pigs/wild boars in each of the groups, and agents (farm/pigs/wild boars) can change states according to schema:

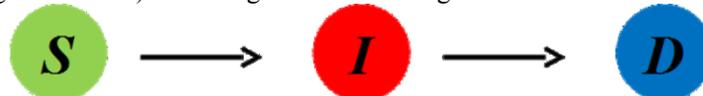

Pic. 2) Mathematical model Susceptible (S) -> Infectious (I) (with rate *inf*). Infectious (I) -> Detected (D) (with rate 1/*durat*)

Let's define important epidemiological variable (Jarynowski, 2011) basic reproduction rate: $R_0 = inf * durat$, where: *inf* – infectivity, *durat* – detection time/elimination time.
The early detection method in zeros approximation can be done by fitting the incidence curve to the exponential function (Diekmann, et al. 2013), resulting estimation of the infectivity coefficient per month (*inf*~1.25) [Fig. 2]. With these calculations and critical value of the basic reproduction rate of the epidemic (R0=1), the effective outbreaks detection and elimination times (D) could be estimated.

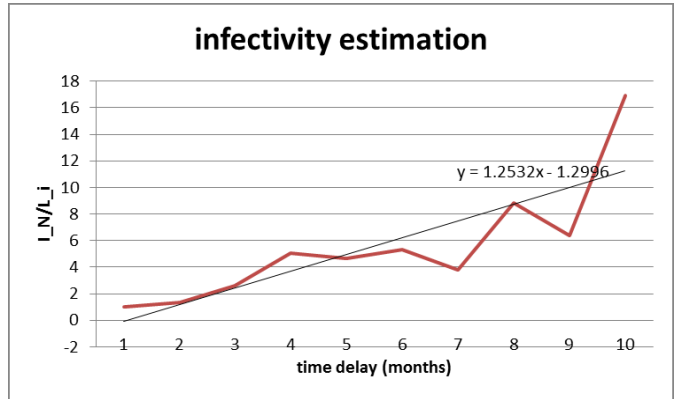

Fig. 2) Preliminary early growth calculations: Incidence~ exp(inf*t) (own calculus)

According to simplified relation in SID (Susceptible, Infectious, Detected) model $R_0=inf*durat$ we observe that detection and elimination time (*durat* ~24 days) is critical to satisfy epidemic condition ($R_0=1$).

**Preliminary Results – Logistic Regressions:**

We cleaned the data registered by Polish Authorities (because of unacceptable amount of errors in registry 2014-2017, as wrong geocoding, etc.). We choose several regression models for each affected counties (18 poviats up to August 2017).

For explained (dependent) variable:
- explained variable is 0/1 (infected/not infected county) [Tab. 2]
- explained variable is integer (number of events understood as sum of outbreaks and cases in county) [Tab. 3]
- explained variable is time (arrival time of ASF for given county) [Tab. 1]

For explaining variables:
- distance to the Belarus boarder;
- heads of pigs, coverage of forests, human population [Pic. 3].

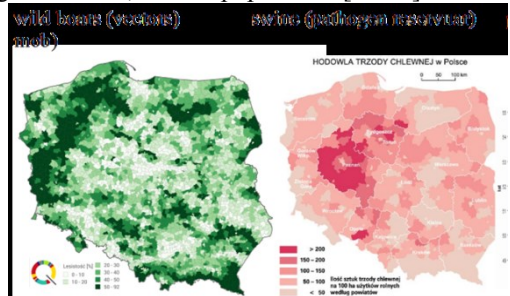

Pic. 3) Data used as predictors in logistic regressions [Left] Forest Coverage, [Right] Amount of pig head [Source www.gus.gov.pl]

TAB. 1 DEPENDENT VARIABLE: ARRIVAL TIME OF INFECTIONS

| Effect | Univariate Tests of Significance for arrival_time | | |
|---|---|---|---|
|  | SS | F | p |
| dist to boarder | 13405334 | 112.68 | 0.02 |
| swine heads | 41404 | 0.35 | 0.56 |
| forest cov | 386987 | 3.25 | 0.09 |
| Error | 1784447 | | |

TAB. 2 DEPENDENT VARIABLE: INFECTION IN A COUNTY (0/1)

infection - Odds Ratios (powiat1)
Distribution : BINOMIAL, Link function: LOGIT
Modeled probability that kod = 1

| Effect | Level of Effect | Column | Odds Ratio | Lower CL 95.0% | Upper CL 95.0% | p |
|---|---|---|---|---|---|---|
| Intercept | | 1 | | | | |
| boarder | | 2 | 0.104497 | 0.039902 | 0.273663 | 0.000004 |
| swine | | 3 | 1.000016 | 1.000007 | 1.000025 | 0.000751 |
| forest | | 4 | 1.030135 | 0.986176 | 1.076053 | 0.182094 |

TAB. 3 DEPENDENT VARIABLE: NUMBER OF CASES IN 2017

| N=11 | Regression Summary for Dependent Variable: number of cases) R= .73085690 R2= .53415180 Adjusted R2= .41768976 F(2,8)=4.5865 p<.04710 Std.Error of estimate: 31.978 | | | | | |
|---|---|---|---|---|---|---|
| | b* | Std.Err. of b* | b | Std.Err. of b | t(8) | p-value |
| Intercept | | | -62.5642 | 40.79639 | -1.53357 | 0.163678 |
| swine | 0.813589 | 0.268631 | 218.6065 | 72.17960 | 3.02865 | 0.016343 |
| forest | 0.353494 | 0.268631 | 1.4202 | 1.07927 | 1.31591 | 0.224660 |

**Network representation of outbreaks:**

For each affected counties (18 poviats up to August 2017) time series of ordered arrival times is construct T=[t1,.., tx,.., t18], where t1=1 (week of introduction to Poland).

We obtain the directed cumulative network [Pic. 4] from 100 trees representing 100 realization of observed outbreak. We cut off links which appears only once.

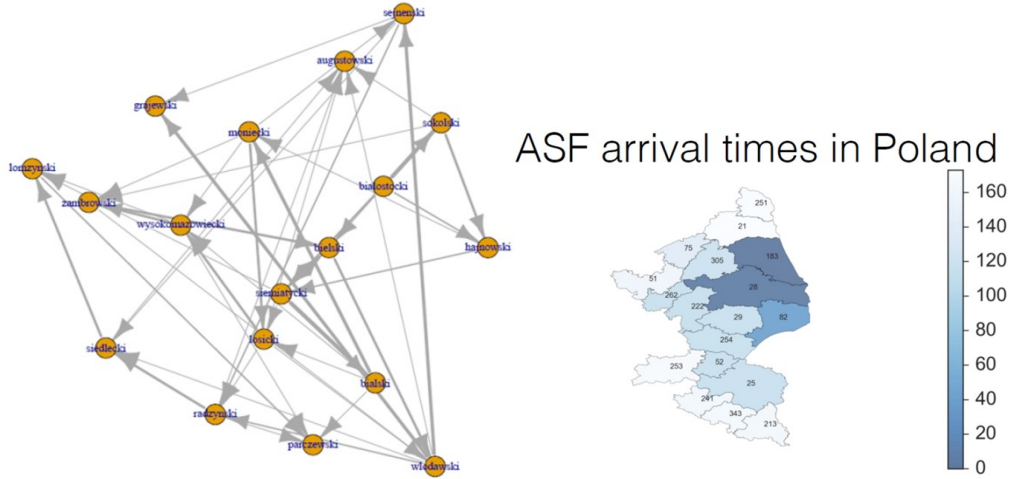

Pic. 4) [Left] Infections paths of counties with only arrival time as predictor. [Right] ASF Arrival times in counties.

**Preliminary Results - Landscape-based propagation:**

In the spatial model we use forest coverage, pig population in poviats and the distance between centroids of poviats. We use pseudo-gravitational models of short and long-range interactions referring to the socio-migratory behavior of wild boars and the pork production chain. We do not estimate directly model parameters and only run over suitable parameter space. We run set of simulations for selected subspace of parameters *a*- swine amount significance, *b*- disease vectors (wild boards) significance, *c*- pork production chain significance.

$$p_{ij} \sim \frac{a(P_i * P_j)}{1+d_{ij}} + \frac{b(F_i * F_j)}{1+d_{ij}^2}, \quad g_{ij} \sim p_{ij} * c$$

where
*a*, *b*, *c* – simulation parameters;
*i*, *j* – poviats;
*P* – normalized amount of pigs;
*F* – coverage of forests;
$p_{ij}$ – probability of infection from a neighbor;
$g_{ij}$ – probability of infection from a whole networks;
$d_{ij}$ – angular distance between centroids of poviats.

We use the simplest SI (Susceptible, Infectious) model on the level on poviats, where with probability $p_{ij}$ disease can be transmitted from already affected poviat *i* to an previously not affected poviat *j*, which is a neighbor to *j*. Long distance transmission can be done with probability $g_{ij}$.

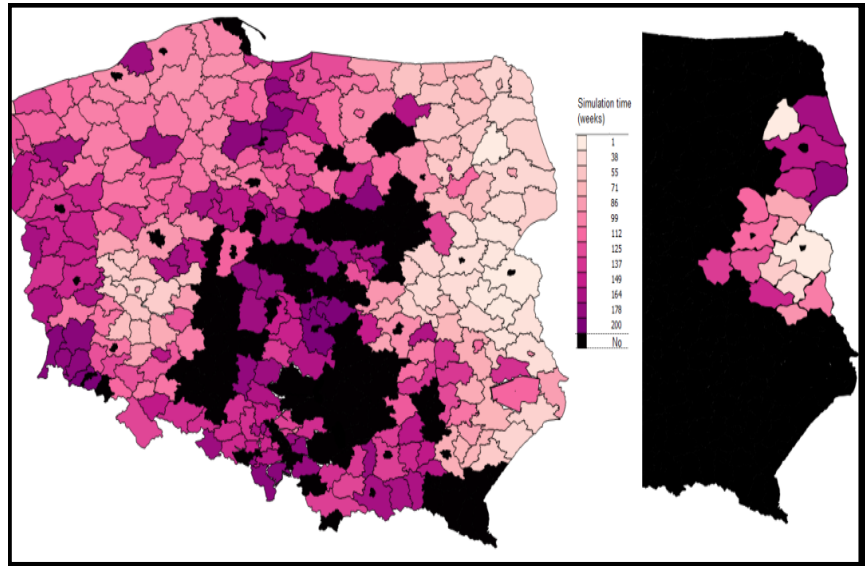

Pic. 5) Examples of simulation spread with seeds in poviats Mońki and Biała Podlaska for next 4 years with possible propagation barriers around invaded infected regions. [Left] Medium swine and vector significance (*a,b*), high pork production chain factor (*c*). [Right] Low swine and vector significance (*a,b*), no pork production chain factor (*c*).

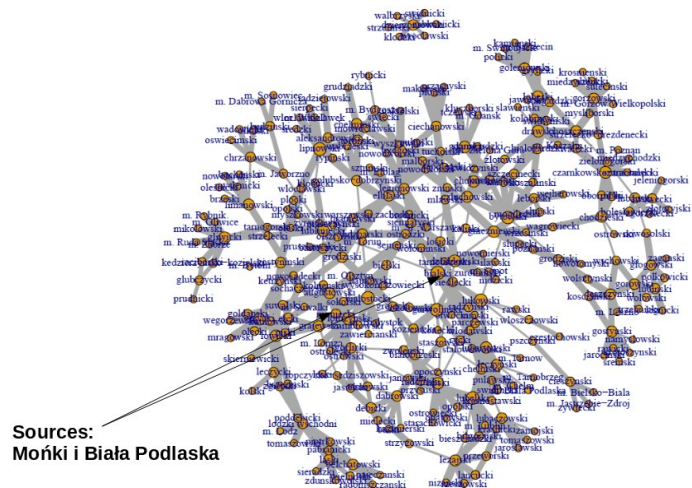

Pic. 6) Most likely paths of infection. Reorganized map of Poland according to ASF risk.

**Preliminary Conclusions:**

ASF not only veterinary public health problem, but probably the most important current economic threat in Eastern Europe causing already few billions EUR yearly losses. It spreads from East to West of Europe with speed of around 200km per year. In future, we want to predict most likely paths of propagation.

From early warning perspective - It is enough to control epidemic with elimination time D shorter than 3 weeks (where 1-2 weeks are the time from exposure to the first clinical symptoms).

The most important conclusion from regressions is, that pig heads strongly explain both probability of introducing ASF as well as epidemic size (which is in contradiction to Estonian study (Depner, et al. 2017)). On the other hand, the forest coverage explains a little the arrival time of ASF (the same as in Estonian study (Depner, et al. 2017)).

Spatial modeling in a certain range of parameters proves the existence of a natural protective barrier within boarders of the 'Congress Poland' [Pic. 5]. The spread of the disease to the 'Greater Poland' should result in the accelerated outbreak of ASF [Pic. 5].

We reconstruct the most likely future paths of infection and classified regions possible for paths as "Kujawy" and direct jumps into "Greater Poland" – Polish Swine district [Pic. 6].

The epidemiological situation in regards to ASF in pigs in Poland seems to be controlled (Pejsak, et al. 2017), but inversely further long-distance ASF spread in wild boar by human difficult to control issue, but easily predictive by regressions models. In future our models would be equipped with decision support systems as a tool for veterinarians.

REFERENCES

Depner, K., Gortazar, C., Guberti, V., Masiulis, M., More, S., Oļševskis, E., ... & Gogin, A. (2017). Epidemiological analyses of African swine fever in the Baltic States and Poland. EFSA Journal, 15(11).

Diekmann O., Heesterbeek, J.A.P. and Britton, T. (2013). Mathematical tools for understanding infectious disease dynamics. Princeton UP.

Jarynowski, A. (2011). Human-Human interaction: epidemiology.in Life-time of correlations, WN: Wroclaw/Glogow.

Mur, L., Martínez-López, B., Martínez-Avilés, M., Costard, S., Wieland, B., Pfeiffer, D. U., & Sánchez-Vizcaíno, J. M. (2012). Quantitative risk assessment for the introduction of African swine fever virus into the European Union by legal import of live pigs. Transboundary and emerging diseases, 59(2), 134-144.

Pejsak, Z., Wozniakowski, G., Smietanka, K., Zietek-Barszcz, A., Bocian, L., Frant, M., & Niemczuk, K. (2017). Przewidywany rozwój sytuacji epizootycznej w zakresie afrykańskiego pomoru świń w Polsce. Życie Weterynaryjne, 92(04).